\documentclass[twocolumn]{article}

\usepackage{amssymb}
\usepackage{sandro}
\usepackage{xspace}

\newcommand{\smallromani}{\renewcommand{\theenumi}{\roman{enumi}}
  \renewcommand{\labelenumi}{(\theenumi)}}

\newcommand{\parentalphi}{\renewcommand{\theenumi}{\alph{enumi}}
  \renewcommand{\labelenumi}{(\theenumi)}}

\newcommand{\la}{\ensuremath{\:\leftarrow\:}}
\newcommand{\ra}{\ensuremath{\:\rightarrow\:}}

\newcommand{\I}{\ensuremath{\:\cap\:}}

\newcommand{\HB}{\hfill{$\square$}}

\renewcommand{\u}{\ensuremath{\:\cup\:}}
\renewcommand{\i}{\ensuremath{\:\cap\:}}
\newcommand{\cont}{\ensuremath{\:\sqsubseteq\:}}
\newcommand{\lhs}{\ensuremath{\lambda}}
\newcommand{\rhs}{\ensuremath{\varrho}}
\newcommand{\der}{\ensuremath{\:\vdash\:}}

\newcommand{\nder}{\ensuremath{\:\not\vdash\:}}
\newcommand{\QQ}{\ensuremath{{\cal Q}}\xspace}
\newcommand{\cc}{\ensuremath{{\cal Q}}\xspace}
\newcommand{\pp}{\ensuremath{{\cal P}}\xspace}
\newcommand{\rr}{\ensuremath{{\cal R}}\xspace}
\newcommand{\gr}{\ensuremath{{\cal G}}\xspace}
\newcommand{\sr}{\ensuremath{{\cal S}}\xspace}

\newcommand{\cred}{\ensuremath{\emph{stmt}}\xspace}

\newcommand{\coa}[2]{\ensuremath{\langle #1,\ #2 \rangle}} %
\newcommand{\cob}[3]{\ensuremath{\langle #1,\ #2,\ #3 \rangle}} %

\newcommand{\semsp}[2]{\ensuremath{[\![#1]\!]_{\mathit{SP(#2)}}}}
\newcommand{\semub}[2]{\ensuremath{[\![#1]\!]_{\emph{UB}(#2)}}}
\newcommand{\semlb}[2]{\ensuremath{[\![#1]\!]_{\emph{LB}(#2)}}}

\newcommand{\g}[2]{\ensuremath{\mathit{\Gamma}_{#2}(#1)}}
\newcommand{\ggr}[3]{\ensuremath{\mathit{\Gamma}_{#2}^{#3}(#1)}}
\newcommand{\head}{\ensuremath{\mathit{head}}}

\newcommand{\cs}{\ensuremath{\mathit{CurrentSet}}}
\newcommand{\os}{\ensuremath{\mathit{OldSet}}}
\newcommand{\jssymb}{\ensuremath{{\cal J\!S}}}
\newcommand{\js}[2]{\ensuremath{{\jssymb}_{#2}(#1)}}

\newcommand{\Ground}{\ensuremath{\mathit{Ground}}}
\newcommand{\mytp}{\ensuremath{T_P}}
\newcommand{\mytpn}[1]{\ensuremath{T_P\!\uparrow^{#1}}}
\newcommand{\mytpp}{\ensuremath{T_{P'}}}
\newcommand{\mytppn}[1]{\ensuremath{T_{P'}\!\uparrow^{#1}}}
\newcommand{\mytpsn}[2]{\ensuremath{T_{\SP{#1}}\!\uparrow^{#2}}}

\newcommand{\includes}{\ensuremath{\longleftarrow}}
\newcommand{\Cred}[2]{\ensuremath{#1\!\includes\! #2}}
\newcommand{\excred}[2]{\mbox{#1 $\longleftarrow$ #2}}

\newcommand\comment[1]{}
\newcommand\discuss[1]{}
\newcommand\todo[1]{}

\newcommand{\sfskip}{\vspace*{0.06in}}

\newcommand\PP{\ensuremath{{\cal P}}\xspace}
\newcommand{\SP}[1]{\ensuremath{\mathit{SP(#1)}}\xspace}

\newcommand{\UB}[1]{\ensuremath{\mathit{UB(#1)}}\xspace}

\newcommand{\ub}{\ensuremath{\mathit{ub}}\xspace}
\newcommand{\lb}{\ensuremath{\mathit{lb}}\xspace}

\newcommand{\lif}{\ensuremath{:\!\!-}\;}

\newcommand{\minstate}[2]{\ensuremath{\pp|_\rr}\xspace}

\newcommand{\ie}{{i.e.}\xspace}

\newcommand{\coNEXP}{\ensuremath{\mathbf{coNEXP}}\xspace}

\newcommand\SRT{\ensuremath{RT_0}\xspace}

\newcommand{\RoleNames}{\ensuremath{\sf Names(\PP)}\xspace}
\newcommand{\Principals}{\ensuremath{\sf Principals(\PP)}\xspace}
\newcommand{\Roles}{\ensuremath{\sf Roles(\pp)}\xspace}

\newcommand{\Emergency}{\ensuremath{\mathit Emergency}\xspace}
\newcommand{\hazmatPersonnel}{\ensuremath{\mathit hazmatPersonnel}\xspace}
\newcommand{\ATF}{\ensuremath{\mathit ATF}\xspace}
\newcommand{\hazmatDB}{\ensuremath{\mathit hazmatDB}\xspace}
\newcommand{\Police}{\ensuremath{\mathit Police}\xspace}
\newcommand{\Fire}{\ensuremath{\mathit Fire}\xspace}
\newcommand{\responsePersonnel}{\ensuremath{\mathit responsePersonnel}\xspace}
\newcommand{\dept}{\ensuremath{\mathit dept}\xspace}
\newcommand{\hazmatTraining}{\ensuremath{\mathit hazmatTraining}\xspace}
\newcommand{\Rollins}{\ensuremath{\mathit Rollins}\xspace}
\newcommand{\Burke}{\ensuremath{\mathit Burke}\xspace}
\newcommand{\OConnel}{\ensuremath{\mathit O'Connel}\xspace}

\begin{document}

\author{Sandro Etalle\thanks{University of Twente,
       P.O.Box 217,
       7500AE Enschede, The Netherlands
       \texttt{s.etalle@utwente.nl}}
\hspace{10mm}
William H. Winsborough\thanks{
       George Mason University,
       4400 University Drive, MS 4A4,
       Fairfax, VA 22030, USA,
       \texttt{wwinsborough@acm.org}}}

\title{Integrity Constraints in Trust Management\footnote{An extended abstract of this work (without appendix) has appeared in \cite{EW05}. This work was partially supported by the BSIK Freeband project I-Share.}}

\todo{add I-share? - url techrep}
\maketitle
\begin{abstract}
We introduce the use, monitoring, and enforcement of integrity
constraints in trust management-style authorization systems.  We
consider what portions of the policy state must be monitored to detect
violations of integrity constraints.  Then we address the fact that
not all participants in a trust management system can be trusted to
assist in such monitoring, and show how many integrity constraints can
be monitored in a conservative manner so that trusted participants
detect and report if the system enters a policy state from which
evolution in unmonitored portions of the policy could lead to a
constraint violation.
\end{abstract}

\section{Introduction}

\newcommand{\FAB}{\textit{FAB}}
\newcommand{\expert}{\textit{expert}}
\newcommand{\university}{\textit{university}}
\newcommand{\phd}{\textit{phd}}

Trust management~\cite{BFL96} (TM) is an approach to managing
authorization in environments where authority emanates from multiple
sources.  Authorization policy consists of statements issued by many
participants, and resource sharing is facilitated by delegating
authority from one principal to another.

A particular authorization is decided by posing a query to the system.
An evaluation procedure combines the statements issued by all relevant
principals to derive the query's answer.  By adding or removing a
policy statement, a principal can potentially affect many
authorizations of many principals.

One of the difficulties of operating in such a context is that at
present no system exists for monitoring unexpected consequences of
policy changes made by other principals. Basically, in present TM systems,
delegating trust implies losing a great deal of control on the policy
involved the delegation. Let us first see three example of this.

Firstly, resources may become unavailable unexpectedly. Consider for
instance a team leader who needs to be informed if members of his team
suffer interruption in their authorization for mission-critical
resources.  If the team's mission involves rapid response, the
notification of interruption should not depend on team members
attempting to access a critical resource and discovering its
unavailability only because the attempt fails.  What is needed is that
the policy change triggers a procedure that pushes the notification to
the team leader.

Secondly: properties such as mutual exclusion cannot be guaranteed. While in the above
example, the exceptional state involved someone losing authorization,
Having someone unexpectedly gain authorization can be just as
important to detect. For instance, it should be possible to trigger an
action if a principal becomes authorized for two mutually exclusive
purposes.  Mutual exclusion is an approach often used, for instance in
RBAC systems~\cite{SCFY96}, to enforce separation of duty, a classic
device aimed at preventing fraud.  By ensuring that no individual is
authorized to complete all parts of a sensitive task, the technique
ensures that only a colluding group could misuse the capability.
Because the participants in a trust management system are autonomous,
it is in general not possible to prevent a principal being given two
authorizations.  However, cooperating principals should be able to
prevent another principal from gaining two mutually exclusive
authorizations under the control of the cooperating group.  What is
needed is a way to distribute the mutual exclusivity requirement and
monitor policy evolution to ensure that control over the key
authorizations is not delegated outside the cooperating group.

Thirdly: quality cannot be monitored. Consider the situation in which
the principal $A$ states, for instance, that he considers expert
anyone that $B$ considers an expert ($A$ delegates to $B$ the
definition of ``expert''). In addition, $A$ expects experts to have a
PhD degree. Now, $A$ has no way of controlling that all experts added
by $B$ actually have doctorates. Of course, $A$ could modify his
policy as follows ``$A$ considers expert anyone \emph{holding a PhD}
that $B$ considers an expert''. However often it would be preferable
for $A$ to know whether a non-PhD had been added to the expert list
because it might suggest to $A$ that an exception to $A$'s policy is
acceptable, or that some other evolution of $A$'s policy should take
place (perhaps it is time to revoke the trust in $B$'s experts).
Thus, what $A$ needs is to be able to monitor whether $B$ ever decides
that a non-PhD is an expert. Notice that this is what would happen in
practice: before delegating to $B$ the definition of expert $A$ would
normally put in place a monitoring activity to guarantee that $B$'s
expert fulfill the quality criteria. Unfortunately, present decentralized TM
systems do not allow for such monitoring.

Summarizing, there is a need for a mechanism to monitor a TM system
and to reveal when an exceptional state has been entered so that
appropriate steps can be taken proactively. Ideally, it would even be
possible to enlist the assistance of others in preventing exceptional
states from arising.  The problem of providing such a monitoring
system is aggravated by the fact that changes are made by autonomous
principals that may not agree or be trusted to assist in the
monitoring.

In this paper we introduce a new trust management construct called a
constraint, inspired by integrity constraints in database management
systems (see, e.g. \cite{GGGM98,Das92}), that provides system
participants the ability to monitor the evolution of the policy.  The
author of a constraint receives notification when the constraint is
violated.  This is achieved by enlisting the assistance of principals
to which authority is delegated and triggering constraint checks when
those principals make relevant policy changes.  The emphasis in this
paper is on determining whether a policy change is relevant, or can be
ignored.

In addition we also consider the setting in which some principals are not trusted or willing to help monitoring a constraint. As mentioned above, in some environments, it is not appropriate to
assume that all principals to whom one delegates authority will assist
in monitoring one's constraints.  By providing a sufficiently
expressive constraint language, we show how to limit to an arbitrary,
specified set those principals that are trusted to cooperate in
monitoring a constraint.  This is done by allowing a constraint to
express a security analysis problem of the kind formulated by Li et
al.~\cite{LWM03}.  Such a constraint quantifies over policy states
that are reachable by policy changes made by untrusted principals
asking whether a given query holds either in all reachable states
(universal quantification) or in some reachable state (existential
quantification).  By checking such a constraint each time the trusted
principals make relevant policy changes, and committing their changes
only if the constraint is satisfied, the trusted principals can ensure
that a state violating the constraint is never entered, no matter what the untrusted principals do.   They are able
to do this because the untrusted principals are unable to affect the
validity of the constraint.

The technical contribution in this paper is a method to identify portions of the policy
state that must be monitored in order to detect constraint violations.
We do this first under the assumption that all principals in the
system can be trusted to assist in monitoring the portion of the
policy state under their control.  We then relax this assumption by
requiring only that a given portion of the policy can be reliably
monitored.  In this case, monitoring is carried out by using security
analysis to assess the possibility of the constraint becoming violated
by policy changes that cannot be monitored directly.

Section~\ref{sec:preliminaries} discusses the TM policy language that
we use.  Section~\ref{sec:main} identifies the portion of the policy
state to be monitored for constraint violations, assuming all portions
can be monitored.  Section~\ref{sec:safety} shows how to monitor
constraints for potential violations when not all parts of the policy
state can be monitored directly.  Section~\ref{sec:relatedwork}
discusses related work.  Section~\ref{sec:conclusion} concludes.  
Some proofs are reported in the appendix.

\section{Preliminaries}
\label{sec:preliminaries}

Trust
management~\cite{BFL96,RFC2704,BFIK99,RL96,RFC2693,CEEFMR01,GJ00,Jim01,LGF03,LWM03,LMW02,LM03,Wee01}
is an approach to access control in decentralized distributed systems
with access control decisions based on policy statements issued by
multiple principals.  In trust management systems, statements that are
maintained in a distributed manner are often digitally signed to
ensure their authenticity and integrity; such statements are sometimes
called \emph{credentials} or \emph{certificates}.  This section
presents the trust management language \SRT\ \cite{LWM03}, which we
use in this paper.

\subsubsection*{The Language \SRT}
A \emph{principal} is a uniquely identified individual or
process. Principals are denoted by names starting with an uppercase,
typically, $A$, $B$, $D$.

A principal can define a \emph{role}, which is indicated by
principal's name followed by the \emph{role name}, separated by a
dot. For instance $A.r$, and $\mathit{GMU.students}$ are roles. For
the sake of simplicity we assume that $A$ is the
\emph{owner} (or the administrator) of $A.r$, 
though the results of this papers apply also in the case $A.r$ is
owned by some other principal. We use names starting with a lowercase
letter (sometimes with subscripts) to indicate role names.

A role denotes a set of principals (the principals that populate it,
i.e., the members of the role). To indicate which principals populate
a role,
\SRT\ allows the owning principal to 
issue four kind of \emph{policy statements}:

\begin{itemize}
   \item \emph{Simple Member}: $A.r\includes D$

With this statement $A$ asserts that $D$ is a member of $A.r$.

 \item \emph{Simple Inclusion}: $A.r\includes B.r_1$

With this statement $A$ asserts that $A.r$ includes (all members of)
$B.r_1$.  This represents a delegation from $A$ to $B$, as $B$ may add
principals to become members of the role $A.r$ by issuing statements
defining (and extending) $B.r_1$.

   \item \emph{Linking Inclusion}: $A.r\includes
  A.r_1.r_2$

We call $A.r_1.r_2$ a \emph{linked role}.  With this statement $A$
asserts that $A.r$ includes $B.r_2$ for every $B$ that is a member of
$A.r_1$.  This represents a delegation from $A$ to all the members of
the role $A.r_1$.

   \item \emph{Intersection
  Inclusion}:  $A.r\includes B_1.r_1\cap B_2.r_2$

We call $B_1.r_1\cap B_2.r_2$ an \emph{intersection}. With this
statement $A$ asserts that $A.r$ includes every principal who is a
member of both $B_1.r_1$ and $B_2.r_2$.  This represents partial
delegations from $A$ to $B_1$ and to $B_2$.
\end{itemize}
For any statement \Cred{A.r}{e}, $A.r$ is called the \emph{head} and
$e$ is called the \emph{body} of the statement.  We write
$\head(\Cred{A.r}{e}) = A.r$.  The set of statements having head $A.r$
is called the \emph{definition} of $A.r$.

The definition of \SRT given here is a slightly simplified (yet
expressively equivalent) version of the one given in~\cite{LWM03}.  A
\emph{policy state} (\emph{state} for short, indicated by \PP) 
is a set of policy statements.  Given a state \PP, we define the
following: $\Principals$ is the set of principals in \PP, \RoleNames
is the set of role names in
\PP, and $\Roles=\{A.r \mid A\in
\Principals, r\in \RoleNames\}$.

To express constraints, we need one last definition: 

\begin{definition} \emph{Positive roles expressions} 
are defined by the following grammar:
\begin{itemize}
\item sets of principals are positive role expressions,
\item roles are positive role expressions,
\item union and intersections of positive role expressions are
 positive role expressions.
\HB
\end{itemize}
\end{definition}
E.g., $A.r$, $A.r \:\cup\: \{A,B\}$ and $A.r \:\cap B.r_1.r_2$.
Positive role expressions, and are denoted by Greek letters, $\phi,
\lambda$, and $\rho$.  A positive role expression containing no roles
(but only sets of principals) is called \emph{static}.

\subsubsection*{Semantics}
The semantics of a policy state is defined by translating it into a
logic program. The \emph{semantic program}, $\SP{\PP}$, of a state
\PP, is a Prolog program has one ternary predicate $m$. Intuitively,
$m(A,r,D)$ means that $D$ is a member of the role $A.r$.

\begin{definition}[Semantic Program] \rm \label{def:sp}
\sloppypar
Given a state \PP, the \emph{semantic program} $\SP{\PP}$ for it is
the logic program defined as follows: (here symbols that start with
``$?$'' represent logical variables)
\begin{itemize}
\item For each $\Cred{A.r}{D} \in \PP$ add to $\SP{\PP}$ the clause\\ 
$m(A, r, D)$
\item For each $\Cred{A.r}{B.r_1} \in \PP$, add to $\SP{\PP}$ the clause\\
$m(A, r, ?Z) \lif m(B, r_1, ?Z)$
\item For each $\Cred{A.r}{A.r_1.r_2} \in \PP$ add to $\SP{\PP}$ the clause\\
$m(A, r, ?Z) \lif m(A, r_1, ?Y),\, m(?Y, r_2, ?Z)$
\item
For each $\Cred{A.r}{B_1.r_1 \cap B_2.r_2} \in
\PP$ add to $\SP{\PP}$ the clause\\
$m(A, r, ?Z) \lif m(B_1, r_1, ?Z),\, m(B_2, r_2, ?Z).$
\HB
\end{itemize}
\end{definition}
We can now define the semantics of a role in a state.

\begin{definition}[Semantics] Given a state \PP,
the semantics of a role $A.r$ is defined in terms of atoms entailed by
the semantic program:
\begin{itemize}
\item
$\semsp{A.r}{\PP} = \{Z | \SP{\PP} \models m(A,r,Z)\}$
\HB
\end{itemize}
\end{definition}
We extend this semantics to positive role expressions in the natural
way as follows: 
\begin{eqnarray*}
\semsp{\{D_1,\ldots,D_n\}}{\PP} &=&
\{D_1,\ldots,D_n\}
\\
\semsp{\phi_1 \cup \phi_2}{\PP} &=&
\semsp{\phi_1}{\PP} \cup
\semsp{\phi_2}{\PP}
\\
\semsp{\phi_1 \cap \phi_2}{\PP} &=& \semsp{\phi_1}{\PP} \cap
\semsp{\phi_2}{\PP}
\end{eqnarray*}

\todo{Discuss evaluation briefly.  Move intro of Tp to here.}

\section{Constraints}
\label{sec:main}

Consider a state \pp, which might change in time. We are interested in
defining a \emph{constraint}, which intuitively is a
\emph{query} that is intended to hold throughout the state changes.  To
this end, we focus on the class of constraints already considered for
the purposes of security analysis in~\cite{LMW04}.  These constraints
express set containment.  

\begin{definition}
  A \emph{constraint} is an expression of the form $\coa{O}{\lhs
  \cont \rhs}$, in which $O$ is a principal called the
  \emph{owner} of the constraint, and
  \lhs\ and \rhs\ are positive role expressions.
\HB
\end{definition}

The following definition clarifies that $\cont$ represents set
containment.

\begin{definition}
  Let $\pp$ be a state and ${\cc}$ be the constraint
  $\coa{O}{\lhs \cont \rhs}$, we say that
\begin{itemize}
\item
$\pp$ \emph{satisfies} $\cc$ (${\pp} \der {\cc})$ iff 
$\semsp{\lhs}{\pp} \subseteq \semsp{\rhs}{\pp}$ 
\end{itemize}
(${\pp}$ \emph{violates} \cc\ otherwise)
\HB
\end{definition}

Constraints of this form can capture many important and intuitive
requirements.

\begin{itemize}
\item Consider \coa{O}{\{\mbox{Bob}\} \i A.r \sqsubseteq \emptyset}.
  This constraint captures a safety requirement that Bob must not
  become a member of $A.r$.
\item The constraint \coa{O}{\{\mbox{Alice}\} \cont A.r} captures the
  availability requirement that Alice must be authorized for $A.r$.
\item \sloppypar
The constraint \coa{O}{A.\mbox{manager} \i B.\mbox{controller}
    \cont \emptyset} captures the mutual exclusivity requirement that
  no one must be authorized for both $A.$manager and $B.$controller.
\end{itemize}

\begin{table*}
\centering
\caption{Policy State of Example \ref{ex:HAZMAT}}
\begin{tabular}{lr}
\\
\excred{\ATF.\hazmatDB}{\Rollins}     & (1)
\\
\excred{\Emergency.\hazmatPersonnel}{\Emergency.\responsePersonnel
 $\cap$ \ATF.\hazmatTraining} & (2)
\\
  \excred{\Emergency.\responsePersonnel}{\Emergency.\dept.\responsePersonnel}
 & (3)
\\
  \excred{\Emergency.\dept}{\Fire} & (4)
\\
  \excred{\Emergency.\dept}{\Police} & (5)
\\
  \excred{\ATF.\hazmatTraining}{\Rollins} & (6)
\\
  \excred{\ATF.\hazmatTraining}{\Burke} & (7)
\\
  \excred{\ATF.\hazmatTraining}{\OConnel} & (8)
\\[2mm]
\mbox{Additional Statements}
\\[2mm]
 \excred{\Police.\responsePersonnel}{\Rollins}     & (9)
\\
 \excred{\Police.\responsePersonnel}{\Burke}     & (10)
\\[2mm]
\mbox{The semantics of } \pp = \{(1), \ldots, (8)\} \mbox{ is}\\[2mm]
$\begin{array}{rcl}
\semsp{\ATF.\hazmatDB}{\pp} & =  & \{\Rollins\}\\
\semsp{\ATF.\hazmatTraining}{\pp} & =  & \{\Rollins, \Burke, \OConnel \}\\
\semsp{\Emergency.\hazmatPersonnel}{\pp} & =  & \emptyset \\
\semsp{\Emergency.\responsePersonnel}{\pp} & =  & \emptyset \\
\semsp{\Emergency.\dept}{\pp} & =  & \{\Fire, \Police \} \\[2mm]
\end{array}$\\
\mbox{The semantics of } \pp' = \pp \u \{(9), (10)\} \mbox{ is}\\[2mm]
$\begin{array}{rcl}
\semsp{\ATF.\hazmatDB}{\pp'} & =  & \{\Rollins\}\\
\semsp{\ATF.\hazmatTraining}{\pp'} & =  & \{\Rollins, \Burke, \OConnel \}\\
\semsp{\Emergency.\hazmatPersonnel}{\pp'} & =  & \{\Rollins, \Burke \} \\
\semsp{\Emergency.\responsePersonnel}{\pp'} & =  & \{\Rollins, \Burke \} \\
\semsp{\Emergency.\dept}{\pp} & =  & \{\Fire, \Police \} \\
\semsp{\Police.\responsePersonnel}{\pp'} & =  & \{\Rollins, \Burke \} \\
\end{array}$
\end{tabular}
\end{table*}

\begin{example} \label{ex:HAZMAT}
\sloppypar
  Suppose the Bureau of Alcohol, Tobacco, Firearms and Explosives
  (\ATF) operates a database containing information about hazardous
  materials (HAZMAT) for use by emergency response personnel.  The
  \ATF individually authorizes users so as to retain tight control
  over the sensitive information contained in the database.  It does
  this by issuing statements such as:

\sfskip
\noindent
\begin{tabular}{lr}
\excred{\ATF.\hazmatDB}{\Rollins}     & (1)
\end{tabular}
\sfskip

\sloppypar
The Emergency Response Center (\Emergency) wants to ensure that all its
hazmat emergency response personnel have access to the database at all
times.  This is expressed by the constraint
\[
\begin{array}{l}
\coa{\Emergency}{\\ 
\ \ \Emergency.\hazmatPersonnel \sqsubseteq
\ATF.\hazmatDB}
\end{array}
\]We assume that \Emergency.\hazmatPersonnel is defined by the
collection of statements $(2) \cdots (8)$ in Table~1.  Suppose the
following two statements are added:

\sfskip
\noindent
\begin{tabular}{lr}
 \excred{\Police.\responsePersonnel}{\Rollins}     & (9)
\\
 \excred{\Police.\responsePersonnel}{\Burke}     & (10)
\end{tabular}
\sfskip

\noindent
When these statements are added, it must be checked whether they cause
violations of the constraint.  Credential (9) does not cause a
violation, but (10) does, and the Emergency Response Center must be
notified accordingly.
\HB
\end{example}

\subsection{Monitoring Constraints}

We now see how we can put in place a system for monitoring constraint
violations.  Let \pp be a state, and consider the constraint $\cc =
\coa{O}{\lhs \cont \rhs}$. Assuming that \pp changes in time, we are interested
in monitoring when \cc is violated.  
\begin{definition} Let ${\pp} \longmapsto {\pp'}$ be a state change
from $\pp$ to $\pp'$. We say that
\begin{itemize}
\item the change \emph{violates} $\cc$ if ${\pp} \der {\cc}$ and
${\pp'} \nder {\cc}$
\end{itemize}
\end{definition}
Notice that if a change violates the constraint, then there exists $D$
such that $D \not\in \semsp{\lhs}{\pp} \setminus \semsp{\rhs}{\pp}$,
while $D \in
\semsp{\lhs}{\pp'} \setminus \semsp{\rhs}{\pp'}$.
This remark points out an important feature of containment
constraints: that if they are violated then there exists a specific
set of principals violating it.

To monitor the system, a feature of $RT$ we are going to exploit is
its monotonicity: adding a statement to $\pp$ cannot cause the set
semantics of a role to shrink.  Similarly, removing a statement cannot
cause the set semantics to grow.  Formally, for each role $A.r$ and
each statement
\cred:
\begin{equation}
\label{eq:monotonicity}
\begin{array}{rcl}
\semsp{A.r}{\pp} &\subseteq& \semsp{A.r}{\pp \u \{\cred \}}\\
\semsp{A.r}{\pp} &\supseteq& \semsp{A.r}{\pp \setminus \{\cred \}} 
\end{array}
\end{equation}
Therefore, adding a statement to \pp can only augment the set
\semsp{\lhs}{\pp} and
\semsp{\rhs}{\pp}. Consequently, if we assume that \pp initially
satisfies $\lhs \cont \rhs$, we see the following:
\begin{itemize}
\item \emph{Adding} a statement to \pp
can yield to a violation of $\lhs \cont \rhs$ only if the addition
affects
\semsp{\lhs}{\pp}.
\item Removing a statement from \pp can yield to a
violation of $\lhs \cont \rhs$ only if the removal affects
$\semsp{\rhs}{\pp}$.
\end{itemize}

We now want to further isolate the roles that might influence the
satisfaction of a constraint.

\begin{example}
\label{exa:bastard}
Consider the following set of statements.
\begin{eqnarray}
\label{exa:ararr}
A.r & \includes& A.r.r\\
\label{exa:arb}
A.r & \includes& B \\
\label{exa:brc}
B.r & \includes&  C \\
C.r & \includes&  D.r\\
E.r & \includes & F
\end{eqnarray}
It is easy to see that $\semsp{A.r}{\pp}$ is $\{ B, C \}$.  Notice now
that if we add a statement $D.r \includes E$, then $\semsp{A.r}{\pp}$
\emph{grows to} $\{B, C, E, F \}$.  Therefore we can say that $D.r$
may \emph{positively} affect $A.r$.  We see that $\{A.r, B.r, C.r, D.r
\}$ is the set of roles that can positively affect $A.r$. Dually, we
can define the set of roles that may affect the \emph{shrinking} of
$\semsp{A.r}{\pp}$. Here, it is easy to see that the only way of
``reducing'' the semantics $\semsp{A.r}{\pp}$ of $A.r$ is by removing
one of the statements (\ref{exa:ararr}), (\ref{exa:arb}) or
(\ref{exa:brc}).  Since these statements define the roles $A.r$ and
$B.r$ we can say that $\{ A.r, B.r \}$ is the set of roles that can
\emph{negatively} affect $A.r$.
\HB
\end{example}

This section constructs two sets of roles whose definitions determine
the membership of a given role $X.u$ in state
\PP.  
If the membership of $X.u$ were to grow, some role in one of these
sets would have to have a new statement in its definition, and if the
membership of $X.u$ were to shrink, some role in the other set would
have to have a statement in its definition revoked.

\subsubsection*{Positive Dependencies} 
Given a set \pp and a role $A.r$ we want to isolate a set
$\g{A.r}{\pp}$ of roles we have to monitor, as they might affect the
\emph{growth} of \semsp{A.r}{\pp}.

\begin{definition} 
\label{def:gp}
Let $A.r$ be a role and \pp be a state;
\g{A.r}{\pp} is the least set of roles containing $A.r$ and
satisfying the following:
\begin{itemize}
\item
If $B.r_0 \in \g{A.r}{\pp}$ and $\Cred{B.r_0}{B.r_1} \in \PP$, then $B.r_1 \in \g{A.r}{\pp}$.
\item
If $B.r_0 \in \g{A.r}{\pp}$ and $\Cred{B.r_0}{B.r_1.r_2} \in \PP$, then $B.r_1 \in \g{A.r}{\pp}$
and $X.r_2 \in \g{A.r}{\pp}$ for all $X \in [\![B.r_1]\!]_{\SP{\PP}}$.
\item If $B.r_0 \in \g{A.r}{\pp}$ and $\Cred{B.r_0}{B_1.r_1 \i \ldots \i
    B_n.r_n} \in \PP$, then for each $i \in [1,n]$ $B_i.r_i \in
  \g{A.r}{\pp}$.\HB
\end{itemize}
\end{definition}

The main properties of \g{.}{\pp}\ we will make use of are summarized
in the following lemma, which is proved in 
the appendix

\begin{lemma} 
\label{lem:gprop}
Let $\pp' = \pp \u \{\cred\}$, where $\head(\cred) \not\in \g{A.r}{\pp}$, then
\begin{enumerate}
\parentalphi
\item $\semsp{A.r}{\pp}  = \semsp{A.r}{\pp'}$, and 
\item $\g{A.r}{\pp} = \g{A.r}{\pp'}$.
\end{enumerate}
Moreover, if $\pp'$ is obtained from \pp by (a) adding zero or more
statements whose head is not in \g{A.r}{\pp}, and (b) removing zero
or more statements, then
\begin{enumerate}
\setcounter{enumi}{2}
\parentalphi
\item $\semsp{A.r}{\pp} \supseteq \semsp{A.r}{\pp'}$, and 
\item $\g{A.r}{\pp} \supseteq \g{A.r}{\pp'}$.
\HB
\end{enumerate}
\end{lemma}

\begin{example}\mbox{}
\label{ex:gamma}
\begin{itemize}
\item \sloppypar Returning to Example \ref{ex:HAZMAT}, the left-hand
  side of the constraint\\ {\small \Emergency.\hazmatPersonnel
  $\sqsubseteq$ \ATF.\hazmatDB} is \Emergency.\hazmatPersonnel.  So
\[
\begin{array}{llll}
\multicolumn{3}{l}{\g{\Emergency.\hazmatPersonnel}{\pp}} = \\
\ \hspace{5mm}\ &\{ & \mbox{\Emergency.\hazmatPersonnel,} \\
&& \mbox{\Emergency.\responsePersonnel,}\\
&&  \mbox{\ATF.\hazmatTraining,} \\
&& \mbox{\Emergency.\dept,} \\
&& \mbox{\Fire.\responsePersonnel,}\\
&& \mbox{\Police.\responsePersonnel} & \}
\end{array}
\]
is the set of roles for which addition of new statements must be
monitored.
\item Consider the policy state in Example \ref{exa:bastard}. Then
  $\g{A.r}{\pp} = \{A.r, B.r, C.r, D.r \}$.
\item 
  Suppose \pp contains only the statement $\{A.r_0 \includes A.r_1.r_2,
  \}$. Then $\g{A.r_0}{\pp} = \{ A.r_0,\ A.r_1 \}$, and $\semsp{A.r_0}{\pp} =
  \emptyset$. Now, if we add a new statement $A.r_1 \includes B$ to \pp 
  (obtaining $\pp'$) then $\semsp{A.r_0}{\pp'}$ is still the empty
  set, while $\g{A.r_0}{\pp}$ is now $\{ A.r_0, A.r_1, B.r_2 \}$.
\HB
\end{itemize}
\end{example}

\todo{following paragraph has to be re-adapted}
For efficiency reasons, we would like $\g{A.r}{\pp}$ to be as small as
possible, while maintaining the properties stated in Lemma
\ref{lem:gprop}. There are two reasons why $\g{A.r}{\pp}$ is
non-minimal: the first reason is that an intersection inclusion can
act as a filter.  For instance, if $\Cred{A.r}{B_1.r_1 \i B_2.r_2} \in
\pp$ and $\semsp{B_1.r_1}{\pp} =
\emptyset$, there is no point in adding  $B_2.r_2$ to $\g{A.r}{\pp}$ as
any change to $B_2.r_2$ will not affect the membership to $A.r$.  The
second reason concerns linked roles: if $\Cred{A.r}{A.r_1.r_2} \in
\pp$ and there exists no role $B.r_2$ such that for some $D$, $D \in
\semsp{B.r_2}{\pp} \setminus \semsp{A.r}{\pp}$, then we 
could avoid adding $A.r_1$, to $\g{A.r}{\pp}$, as any addition to
$B_2.r_2$ would not affect the membership to $A.r$. However, refining
the definition $\g{A.r}{\pp}$ to take these factors into consideration
would make its definition more complex than seems practical.

\subsubsection*{Negative Dependencies}
Now, we need to isolate the dual of $\g{A.r}{\pp}$, i.e., a set of
roles that might cause $\semsp{A.r}{\pp}$ to shrink. To this end, we
say that that $\Sigma$ is a \emph{\pp-support} of $D$ for $A.r$ if the
roles in $\Sigma$ carry enough information to demonstrate that $D \in
\semsp{A.r}{\pp}$.  We denote by $\pp|_{\Sigma}$ the
restriction of $\pp$ to the roles in $\Sigma$, $\pp|_{\Sigma} =
\{\cred \in \pp|\head(\cred) \in \Sigma \}$

\begin{definition}
\label{def:support} 
Let $A.r$ be a role, $D$ be a principal, $\pp$ be a set of statements
and and $\Sigma$ be a set of roles.
\begin{itemize}
\item We say that $\Sigma$ is a \emph{\pp-support of} $D$ \emph{for} $A.r$ if
  $D \in \semsp{A.r}{\pp|_{\Sigma}}$.
\item  For $L\subseteq \Principals$, we say that $\Sigma$ is a
  \emph{\pp-support of} $L$ \emph{for} $A.r$ if
  $D \in \semsp{A.r}{\pp|_{\Sigma}}$ for every $D \in L$.
\item We say that $\Sigma$ is a \emph{\pp-support for} $A.r$ if and
  only if it is a $\pp$-support of every $D \in \semsp{A.r}{\pp}$.
\HB
\end{itemize}
\end{definition}

\begin{example}\mbox{}
\label{exa:support}
\begin{enumerate}\smallromani
\item Consider again the policy state in Example \ref{exa:bastard}. 
 Any set containing $\{ A.r, B.r \}$ as a subset is a support for
 $A.r$.
\item In case of redundancies, minimal support might not be unique.  Consider
\sfskip 

\begin{tabular}{llr}
 \hspace*{0.1in} 
&\Cred{A.r}{B.r} \\
&\Cred{A.r}{C.r} \\
&\Cred{B.r}{F} \\
&\Cred{C.r}{F} 
&\end{tabular}
\sfskip

Here, both $\{A.r, B.r \}$ and $\{ A.r, C.r \}$ are support for $A.r$.
\HB
\end{enumerate}
\end{example}

We can now state the counterpart of Lemma \ref{lem:gprop}.
\begin{lemma} 
\label{lem:sprop}
Let $A.r$ be a role, $D$ be a principal, $\pp$ be a state and $\Sigma$
be a \pp-support of $D$ for $A.r$. Then
\begin{enumerate}
\item $D \in \semsp{A.r}{\pp}$
\end{enumerate}
Moreover, if $\pp'$ is obtained from \pp by (a) removing zero or more
statements whose head is not in $\Sigma$, and (b) adding zero or more
statements, then
\begin{enumerate}
\setcounter{enumi}{1}
\item $\Sigma$ is a $\pp'$-support for $A.r$, and therefore
\item $D \in \semsp{A.r}{\pp'}$
\end{enumerate}
\emph{Proof}. Point 1 follows immediately from the fact that, by
monotonicity, $\semsp{A.r}{\pp} \supseteq
\semsp{A.r}{\pp|_{\Sigma}}$. For points 2 and 3, by the construction
of $\pp'$ we have that $\pp|_{\Sigma} \subseteq \pp'$, so the results
follows from the definition of support and the fact that the semantics
is monotonic.
\HB
\end{lemma}

To build a
$\pp$-support of $D$ for $A.r$ one basically has to collect all the
roles used to prove that $D \in
\semsp{A.r}{\pp}$.  In 
the appendix
we give an algorithm to compute
minimal \pp-support while evaluating role membership.
 
\subsubsection*{Putting Things Together}

We can now prove the result we were aiming at. Suppose we need to
deploy the integrity constraint $\cc = \lhs \cont \rhs$ on $\pp$. The
first step we need to take is to check if \pp satisfies \cc.  This is
can be done as follows:
\begin{enumerate}
\item First, $\semsp{\lhs}{\pp}$ is computed.
\item Then, for each $D \in \semsp{\lhs}{\pp}$, we check that $D \in
  \semsp{\rhs}{\pp}$.
\end{enumerate}
In step 2, while checking that $D \in \semsp{\lhs}{\pp}$ it is usually
possible to build for free a \pp-support of $D$ in $\rhs$.  Once we
have checked that \pp\ satisfies \cc, we want to make sure that
changes to \pp\ do not cause a violation of \cc. For this we have the
following.
\begin{theorem}[Main] 
\label{thm:main}
Assume that $\pp$ satisfies the constraint $\coa{O}{\lhs \cont
\rhs}$. Let $\Sigma$ be a $\pp$-support of $\semsp{\lhs}{\pp}$ for
$\rhs$, and let ${\pp}
\longmapsto {\pp'}$ be a (possibly multistep) change from $\pp$ to $\pp'$. If
\begin{enumerate}
\smallromani
\item $\forall\;\cred \in \pp' \backslash \pp,\ \head(\cred) \not\in \g{\lhs}{\pp}$, and 
\item $\forall\;\cred \in \pp \backslash \pp',\ \head(\cred) \not\in
  \Sigma$
\end{enumerate}
Then ${\pp'}$ satisfies the constraint $\coa{O}{\lhs \cont
\rhs}$ as well.\\[2mm]
\emph{Proof}.\\
\begin{tabular}{rl}
Take any & $D \in \semsp{\lhs}{\pp'}$\\
By Lemma \ref{lem:gprop}, & $D \in \semsp{\lhs}{\pp}$\\
Since by assumption, $\pp \der \lhs \cont \rhs$, & $D \in \semsp{\rhs}{\pp}$\\
By Lemma \ref{lem:sprop}, & $D \in \semsp{\rhs}{\pp'}$
\end{tabular}
\\
Hence the thesis.
\HB
\end{theorem}

Theorem \ref{thm:main} also shows that, as long as the changes to \pp 
satisfy (i) and (ii), we do not have to recompute the set
$\g{\lhs}{\pp}$ or the support $\Sigma$.  Technically, this is due to
the fact that changes satisfying (i) and (ii) do not affect $\Sigma$
(by Lemma \ref{lem:sprop}, $\Sigma$ is still a support of \rhs), and
can only reduce the set $\g{\lhs}{\pp}$ (by Lemma \ref{lem:gprop}).
When statements defining roles in $\g{\lhs}{\pp}$ are issued, (i) is
violated, and when statements defining roles in $\Sigma$ are revoked,
(ii) is violated.  At these times, the constraint must be checked and
the sets $\g{\lhs}{\pp}$ and $\Sigma$ must be recomputed.

The theorem indicates how a system for monitoring constraints should
be deployed: the first step (mentioned above) is to check that \pp
satisfies $\lhs \cont \rhs$. While doing this, we can build an
appropriate $\Sigma$. Secondly, we have to build $\g{\lhs}{\pp}$.
Thirdly, we need to put in place \emph{monitoring} of the roles in
$\Sigma$ and in $\g{\lhs}{\pp}$ such that each time a statement
defining a role in $\g{\lhs}{\pp}$ (resp.\ $\Sigma$) is added to
(resp.\ deleted from) \pp, the constraint owner is warned.  When the
constraint owner receives a warning he has to (a) check whether the
constraint still holds, and (b) recompute $\g{\lhs}{\pp}$ and
$\Sigma$.

\begin{example}\mbox{}
\begin{itemize}
\item
  \sloppypar Returning to Example \ref{ex:HAZMAT}, to monitor
  \coa{\mbox{\Emergency}}{\Emergency.\hazmatPersonnel \sqsubseteq
  \mbox{\ATF.\hazmatDB}}, we must monitor revocation of definitions of
  roles in some \pp-support of each member of
  \semsp{\Emergency.\hazmatPersonnel}{\pp} for \ATF.\hazmatDB.  In
  this example, $\Sigma = \{\ATF.\hazmatDB\}$ is a \pp-support of each
  such member for \ATF.\hazmatDB.  We must also monitor additions to
  \g{\Emergency.\hazmatPersonnel}{\pp}, as discussed in
  Example~\ref{ex:gamma}.  If new statements are added defining other
  roles, no action has to be taken.  Similarly, if statement (10),
  \excred{\Police.\responsePersonnel}{\Burke}, were removed, no action
  would be necessary because \Police.\responsePersonnel is not in
  $\Sigma$.  
\item
Consider now Example \ref{exa:support} (ii), together with the query
$\{ F \} \cont A.r$. To apply Theorem \ref{thm:main}, we have to
choose one support of $F$ for $A.r$ (the two candidate support are
$\{A.r,\ B.r\}$ and $\{A.r,\ C.r\}$) and monitor the roles in it.
Suppose we choose $\Sigma = \{A.r,\ B.r\}$. Suppose we now remove the
statement \Cred{B.r}{F}. This does not yield to a violation of the
constraint, but we do have to recompute $\Sigma$, which now becomes
$\{A.r,\ C.r\}$.  
\item
Finally, it is also instructive to see that
a change in $\g{\lhs}{\pp}$ might require recomputing $\Sigma$, even
if it does not entail a violation of the constraint. Let $\pp$ be the
following set of statements: 
\begin{eqnarray*}
A.r & \includes &E \\
B.r & \includes &C.r\\
B.r & \includes &D.r\\
C.r & \includes &E\\
D.r & \includes &F
\end{eqnarray*}
together with the constraint $A.r \cont B.r$. This constraint is
satisfied and to monitor its evolution we have to monitor the roles in
$\g{A.r}{\pp} = \{A.r\}$ and $\Sigma = \{B.r,\ C.r \}$. Now if we add
the statement $\Cred{A.r}{F}$ then the constraint owner is warned that
a change in $\g{\lhs}{\pp}$ has occurred. The constraint owner can
check that the constraint is still satisfied in $\pp' = \pp \cup
\{\Cred{A.r}{F}\}$; however $\Sigma$ has to be recomputed to take into
account that it should be a $\pp'$-support of $F$ too. The new
$\Sigma$ is $\{B.r,\ C.r,\ D.r \}$.
\HB
\end{itemize}
\end{example}

\subsection{Alternative Support Definition}

We have defined the \emph{\pp-support} $\Sigma$ to be a set of
roles. Alternatively, we could have defined $\Sigma$ to be a set of
\emph{credentials}.

\begin{definition}[Alternative definition of support]
\label{def:altsupport} 
Let $A.r$ be a role, $D$ be a principal, $\pp$ be a set of statements
and and $\Sigma \subseteq \pp$ be a set of credentials
\begin{itemize}
\item We say that $\Sigma$ is a \emph{\pp-support of} $D$ \emph{for} $A.r$ if
  $D \in \semsp{A.r}{\Sigma}$.
\item  For $L\subseteq \Principals$, we say that $\Sigma$ is a
  \emph{\pp-support of} $L$ \emph{for} $A.r$ if
  $D \in \semsp{A.r}{\Sigma}$ for every $D \in L$.
\item We say that $\Sigma$ is a \emph{\pp-support for} $A.r$ if and
  only if it is a $\pp$-support of every $D \in \semsp{A.r}{\pp}$.
\HB
\end{itemize}
\end{definition}

Monitoring constraint using this definition requires more machinery
than using Definition
\ref{def:support}, but it could yield to a more efficient
implementation. With this definition one monitors the
\emph{credentials} and not the \emph{roles} which might 
affect the right hand side of the constraint.  Therefore, to apply
this definition one needs a mechanism for monitoring every single
credential of $\Sigma$ (which might be difficult). 

\begin{theorem}[Main with alternative definition] 
\label{thm:altmain}
\sloppypar
Assume that $\pp$ satisfies the constraint $\coa{O}{\lhs \cont
\rhs}$.
Let $\Sigma$ be a
$\pp$-support of $\semsp{\lhs}{\pp}$ for $\rhs$ (according to
Definition \ref{def:altsupport}), and let ${\pp}
\longmapsto {\pp'}$ be a (possibly multistep) change from $\pp$ to $\pp'$. If
\begin{enumerate}
\smallromani
\item $\forall\;\cred \in \pp' \backslash \pp,\ \head(\cred) \not\in \g{\lhs}{\pp}$, and 
\item $\forall\;\cred \in \pp \backslash \pp',\ \cred \not\in
  \Sigma$
\end{enumerate}
Then ${\pp'}$ satisfies $\coa{O}{\lhs \cont
\rhs}$ as well.
\HB
\end{theorem}

The advantage of Definition \ref{def:altsupport}, is that the
hypothesis of Theorem
\ref{thm:altmain} hold more often than those of Theorem
\ref{thm:main}.  In other words, using Definition \ref{def:altsupport}
one has to check whether the query still holds and to recalculate
$\g{\lhs}{\pp}$ and $\Sigma$ less often than with Definition
\ref{def:support}.

\section{Monitoring When Not All Participants Are Trusted
  to Help}
\label{sec:safety}

The previous section showed how principals in a trust management
system can monitor integrity constraints by monitoring changes in the
definitions of certain roles.  This section considers the problem of
monitoring integrity constraints when not all principals in the system
agree to assist in monitoring their roles.  The idea is to make the
assumption that the owners of a certain set of roles are trusted to
monitor new statements added to their definitions.  We call these the
growth-trusted roles and denote them by \gr.  Similarly, the owners of
a set of shrink-trusted roles, denoted \sr, are trusted to monitor
statements removed from their definitions.  The owners of these roles
are trusted to test whether changes made to untrusted roles could
violate the constraint and, if so, to signal that potential violation.
We call the pair $\rr = (\gr, \sr)$ a \emph{role monitor} because it
indicates the roles that can be monitored with respect to growth and
shrinkage.

\begin{definition}[Reachable]
\label{def:reachable}
In the presence of a role monitor $\rr$, we say that $\pp'$ is
\rr-reachable from $\pp$ if $\pp'$ can be obtained from \pp
without adding any statements defining roles in \gr or removing any
statements defining roles in \sr.  That is to say, $\{\cred \in
\pp' | \head(\cred) \in \gr\} \subseteq \pp$ and $\{\cred
\in \pp | \head(\cred) \in \sr \} \subseteq \pp'$.
\HB
\end{definition}
The problem we address is to monitor whether the system ever enters a
state \pp from which some reachable $\pp'$ violates $\lhs \cont \rhs$.
This problem is closely related to the security analysis
problem~\cite{LMW04}, which also is defined in terms of a role monitor
$\rr = (\gr, \sr)$, although in that context it is called a
restriction rule.  In security analysis, the definitions of roles in
\gr are assumed not to grow and those of roles in \sr, not to shrink;
the security analysis problem is to determine whether other changes to
the policy state could cause a constraint to become violated.  In
\cite{LMW04} it was shown that this problem is decidable (\coNEXP) for
$RT_0$ over the class of constraints we consider here, and that it is
polynomial for an important subclass of those constraints.  What has
not been shown before, and what we show in this section, is how to
identify subsets of \gr and \sr that need to be monitored so that
security analysis can be used to maintain integrity constraints.

In the rest of this section, we introduce alternative semantics that
can be used to answer questions about policy states that are reachable
through changes to the definitions of untrusted roles.  We then
formalize sets of roles that must be monitored and show that
monitoring these roles is sufficient.  Finally, we provide a method
for monitoring integrity constraints when not all principals in the
system are trusted to assist the process.

\subsubsection*{Alternative Semantics}
We now recall two non-standard semantics for a policy state \pp and
role monitor \rr.  These were introduced~\cite{LMW04} for computing
the lower and upper bounds on role memberships under the assumption
that the definition of roles in \gr do not grow and the definition of
roles in \sr do not shrink.  We first recall the lower-bound program
for a state \pp and a restriction \rr; this program enables one to
compute the lower-bounds of every role.

\begin{definition}[Lower-Bound Program \cite{LMW04}] \label{def:lb} \rm
\sloppypar
Given \pp and \rr, the \emph{lower-bound program} for them,
$LB(\pp,\rr)$, is constructed as follows:
\begin{itemize}
\item[(b1)]
For each \Cred{A.r}{D} in $\minstate{\pp}{\rr}$,
 add
\\
 \lb(A, r, D)  
\item[(b2)]
For each \Cred{A.r}{B.r_1} in \minstate{\pp}{\rr},
 add
\\
$\lb(A, r, ?Z) \lif \lb(B, r_1, ?Z)$  
\item[(b3)]
For each \Cred{A.r}{A.r_1.r_2} in \minstate{\pp}{\rr},
 add
\\
  $\lb(A, r, ?Z) \lif \lb(A, r_1, ?Y),\; \lb(?Y, r_2, ?Z)$  
\item[(b4)]
For each \Cred{A.r}{B_1.r_1 \cap B_2.r_2} in \minstate{\pp}{\rr},
 add
\\
  $\lb(A, r, ?Z) \lif \lb(B_1, r_1, ?Z),\; \lb(B_2, r_2, ?Z)$.
\HB
\end{itemize}
\end{definition}

We now recall the upper-bound program for a state \pp and a role
monitor \rr.  This program enables one to simulate the upper-bound of any
role.

\begin{definition}[Upper-Bound Program \cite{LMW04}] \rm \label{def:ub}
Given \pp and $\rr=(\gr,\sr)$, their upper-bound program,
$\UB{\pp,\rr}$, is constructed as follows.  ($\top$ is a special
principal symbol not occurring in \pp, \rr, or any query \QQ.)
\begin{itemize}
\item[(u)] Add $\ub(\top, ?r, ?Z)$
\item[(u0)]
For each $A.r\in \Roles \backslash \gr$, add
\\
 $ \ub(A, r, ?Z) $ 
\item[(u1)]
For each \Cred{A.r}{D} in \pp, add
\\
$\ub(A, r, D) $ 
\item[(u2)]
For each \Cred{A.r}{B.r_1} in \pp, add
\\
 $ \ub(A, r, ?Z) \lif \ub(B, r_1, ?Z)  $ 
\item[(u3)]
For each \Cred{A.r}{A.r_1.r_2} in \pp, add
\\
 $ \ub(A, r, ?Z) \lif \ub(A, r_1, ?Y), \ub(?Y, r_2, ?Z)  $ 
\item[(u4)]
For each \Cred{A.r}{B_1.r_1 \cap B_2.r_2} in \pp,
 add
\\
 $ \ub(A, r, ?Z) \lif \ub(B_1, r_1, ?Z), \ub(B_2, r_2, ?Z)  $ 
\HB
\end{itemize}
\end{definition}

The rules $(u1)$ to $(u4)$ follow from the meanings of the four types
of statements and are similar to the semantic program construction in
Definition~\ref{def:sp}. The rule $(u0)$ means that for any role $A.r$
not in \gr, the upper-bound of $A.r$ contains every principal.  The
rule $(u)$ means that for any role name $r$, the upper-bound of
$\top.r$ contains every principal. This is so because $\top$ does not
appear in $\gr$.  The rule $(u)$ is needed because given
$\Cred{A.r}{A.r_1.r_2}$, where $A.r \in\gr$ and $A.r_1 \not\in\gr$, we
should ensure that the upper-bound of $A.r$ contains every principal.
We define:
\begin{eqnarray}
\semub{A.r}{\pp} &=& \{Z \ | \ \ub(\PP) \models m(A,r,Z)\}\\
\semlb{A.r}{\pp} &=& \{Z \ | \ \lb(\PP) \models m(A,r,Z)\}
\end{eqnarray}
And by definition we have that 
\begin{remark}\mbox{}
\label{rem:semantics}
\begin{itemize}
\item
If $A.r \not\in \sr$ then $\semlb{A.r}{\pp} = \emptyset$. 
\item 
If $A.r \not\in \gr$ then $\semub{A.r}{\pp} = \Principals \u \{ \top \}$. 
\HB
\end{itemize}
\end{remark}

The next theorem gives the link between the two new semantics and the
problem of checking that a constraint is satisfied in all reachable
$\pp'$.

\begin{theorem}[\cite{LMW04}]
\label{thm:strongsemantics}
Let \rr\ be a role monitor, \pp be a state, and $\lhs
\cont \rhs$ be a containment constraint.
\begin{itemize}
\item If $\semub{\lhs}{\pp} \subseteq \semlb{\rhs}{\pp}$ then $\pp' \der
  \lhs \cont \rhs$ for each
  $\pp'$ reachable from \pp,
\item if either $\lhs$ or $\rhs$ is \emph{static} (i.e., it is a set
  of principals) then $\pp' \der \lhs \cont \rhs$ for each $\pp'$
  reachable from \pp implies that $\semub{\lhs}{\pp} \subseteq
  \semlb{\rhs}{\pp}$\footnote{Actually, though we do not prove it
  here, we believe that a stronger version of this part holds, stating
  that if $\g{\lhs}{\pp} \I \g{\rhs}{\pp} = \emptyset$ then $\pp' \der
  \lhs \cont \rhs$ for each $\pp'$ reachable from \pp implies that
  $\semub{\lhs}{\pp} \subseteq
\semlb{\rhs}{\pp}$.}.
\HB
\end{itemize}
\end{theorem}

We now proceed as in the previous section, by identifying the roles we
have to monitor. 

\subsubsection*{Positive Dependencies, with Untrusted Roles}
In the light of Theorem \ref{thm:strongsemantics}, given a state \pp, a
role monitor \rr, and a role $A.r$, we want to isolate a set
$\ggr{A.r}{\pp}{\gr}$ of roles we have to monitor, as they might
affect the
\emph{growth} of \semub{A.r}{\pp}. One might think that when some
roles are untrusted, we need only
restrict $\g{A.r}{\pp}$ to the $\gr$-roles (or to check that
$\g{A.r}{\pp} \subseteq \gr$). The following example shows that this
is not adequate. Consider the constraint $A.r \cont B.r$, where $A.r$
is defined by
\begin{eqnarray}
A.r &\includes &C.r \i D.r\\
D.r& \la& E.r\\
\nonumber
&\ldots
\end{eqnarray}
$A.r$ depends on $C.r$, $D.r$ and $E.r$ (which are in $\g{A.r}{\pp}$),
and, if we used the method of
the previous section, we would have to monitor
all three of them.
We now make two observations about monitoring when it is not possible to
monitor all three roles.
First, if $E.r$ is not in \gr, we cannot monitor it. This implies
that there is no point in monitoring $D.r$ either, as it directly
depends on $E.r$. Second if $D.r$ is not in \gr, there is no point in
monitoring it nor in monitoring $E.r$ (which can only influence $A.r$
via $D.r$).

To cope with this we now define the \pp-core of \gr, which intuitively
contains those role of \gr\ which additionally do not fully depend on
an untrusted role.

\newcommand{\core}[2]{\ensuremath{\mathit{core}_{#2}(#1)}}

\begin{definition}[\pp-Core]
  Let \pp be a state and \gr\ be a set of roles. The
  \emph{\pp-core of} \gr, $\core{\gr}{\pp}$, is the maximal subset of
  \gr such that
\begin{itemize}
\item If $\Cred{A.r}{B.r_1} \in \pp$, and $B.r_1 \not\in \core{\gr}{\pp}$, then $A.r \not\in \core{\gr}{\pp}$
\item
If $\Cred{A.r}{A.r_1.r_2} \in \pp$, and $A.r_1 \not\in \core{\gr}{\pp}$, then $A.r \not\in \core{\gr}{\pp}$.
\item If $\Cred{A.r}{A.r_1.r_2} \in \pp$, and $\exists B \in
  \semub{A.r_1}{\pp}$ such that $B.r_2 \not\in \core{\gr}{\pp}$, then
  $A.r \not\in \core{\gr}{\pp}$.
\item If $\Cred{A.r}{A_1.r_1\cap \ldots \cap A_n.r_n} \in \pp$, and
  for every $i$, $A_i.r_i \not\in \core{\gr}{\pp}$, then $A.r \not\in
  \core{\gr}{\pp}$.
\HB
\end{itemize}
\end{definition}

The following proposition is proved in 
the appendix.

\begin{proposition}
\label{pro:noncore}
Let \pp be a set of statements and \gr\ be a set of roles.  
\begin{itemize}\item
If $A.r
\not\in \core{\gr}{\pp}$, then $\semub{A.r}{\pp} = \Principals \u
\{\top \}$.\HB
\end{itemize}
\end{proposition}

We now construct the set of roles that must be monitored for new
definitions to detect growth in a role's membership.

\begin{definition} 
\label{def:gpgr}
Let $A_0.r_0$ be a role in $\core{\gr}{\pp}$, \rr\ be a role monitor, and
\pp be a state; $\ggr{A_0.r_0}{\pp}{\gr} \subseteq
\Roles$ is the least set
satisfying the following:
\begin{itemize}
\item If $A_0.r_0 \in \core{\gr}{\pp}$, $A_0.r_0 \in \ggr{A_0.r_0}{\pp}{\gr}$.
\item If $A.r \in \ggr{A_0.r_0}{\pp}{\gr}$, and $\Cred{A.r}{B.r_1} \in
  \pp$, then $B.r_1 \in \ggr{A_0.r_0}{\pp}{\gr}$.
\item
If $A.r \in \ggr{A_0.r_0}{\pp}{\gr}$ and $\Cred{A.r}{A.r_1.r_2} \in
\pp$, then $A.r_1 \in \ggr{A_0.r_0}{\pp}{\gr}$
and $X.r_2 \in \ggr{A_0.r_0}{\pp}{\gr}$ for all $X \in \semub{A.r_1}{\pp}$ 
\item If $A.r \in \ggr{A_0.r_0}{\pp}{\gr}$ and $\Cred{A.r}{A_1.r_1
\cap \ldots \cap A_n.r_n} \in \pp$, then,
  for each $i \in [1,n]$ if $A_i.r_i \in \core{\gr}{\pp}$,
  $A_i.r_i \in \ggr{A_0.r_0}{\pp}{\gr}$.
\HB
\end{itemize}
\end{definition}
It is easy to prove by a simple induction on the steps in the
iterative construction of \ggr{A_0.r_0}{\pp}{\gr} that
$\ggr{A_0.r_0}{\pp}{\gr} \subseteq \core{\gr}{\pp}$

We now have the counterpart of Lemma \ref{lem:gprop}.

\begin{lemma} 
\label{lem:gpropsafety}
Assume $\top \not\in \semub{A.r}{\pp}$.
Let $\rr$ be a role monitor, $\pp' = \pp \u \{\cred\}$, where
$\head(\cred) \not\in \ggr{A.r}{\pp}{\gr}$, then
\begin{enumerate}
\parentalphi
\item $\semub{A.r}{\pp}  = \semub{A.r}{\pp'}$, and 
\item $\ggr{A.r}{\pp}{\gr} = \ggr{A.r}{\pp'}{\gr}$.
\end{enumerate}
Moreover, if $\pp'$ is obtained from \pp by (a) adding zero or more
statements whose head is not in \ggr{A.r}{\pp}{\gr}, and (b) removing zero
or more statements, then
\begin{enumerate}
\setcounter{enumi}{2}
\parentalphi
\item $\semub{A.r}{\pp} \supseteq \semub{A.r}{\pp'}$, and 
\item $\ggr{A.r}{\pp}{\gr} \supseteq \ggr{A.r}{\pp'}{\gr}$.
\end{enumerate}
\emph{Proof (sketch).}  The result follows by using reasoning similar
to that used for proving Lemma \ref{lem:gprop}.
\HB
\end{lemma}

\discuss{The above proof would probably be important to the journal
version, as the clauses constructed by UB are quite different from
those constructed by SP. }

\subsubsection*{Negative Dependencies, with Untrusted Roles} 

To handle the right hand side of the constraints we simply have to
generalize Lemma \ref{lem:sprop} in the obvious way by taking into
account the presence of the role monitor. The proof of this lemma is
also identical to that of Lemma \ref{lem:sprop}

\begin{lemma} 
\label{lem:spropsafety}
Let $\rr = (\gr, \sr)$ be a role monitor, $A.r$ be a role, $D$ be a
principal, $\pp$ be a state and $\Sigma$ be a \pp-support of $D$ for
$A.r$ \emph{such that $\Sigma \subseteq \sr$}. Then
\begin{enumerate}
\item $D \in \semlb{A.r}{\pp}$
\end{enumerate}
Moreover, if $\pp'$ is obtained from \pp by (a) removing zero or more
statements whose head is not in $\Sigma$, and (b) adding zero or more
statements, then
\begin{enumerate}
\setcounter{enumi}{1}
\item $\Sigma$ is a $\pp'$-support for $A.r$, and therefore
\item $D \in \semlb{A.r}{\pp'}$.
\HB
\end{enumerate}
\end{lemma}
Recall that by Remark \ref{rem:semantics}, if $A.r \not\in \sr$ then
we have that $\semlb{A.r}{\pp} = \emptyset$.  Consequently, it is easy
to show that if $D \in \semlb{A.r}{\pp}$, then there exists a
\pp-support of $D$ for $A.r$ consisting of roles that are in
$\sr$.

\subsubsection*{Putting Things Together}
We can now prove the result we were aiming at. Differently from the
case in which all roles were trusted, we now want to check that $\lhs
\cont \rhs$ holds in any \rr-reachable state $\pp'$. The additional
problem here is we cannot rely on the cooperation of the roles that
are not in $\gr$ (resp.\ $\sr$) in monitoring the constraint and
telling the constraint owner when a statement defining a role in
$\g{\lhs}{\pp}$ is added (resp.\ a statement defining a role in
$\Sigma$ is removed). Because of this we refer to two ``pessimistic''
semantics, $\semub{\lhs}{\pp}$ and $\semlb{\rhs}{\pp}$, and we check
if $\semub{\lhs}{\pp} \subseteq \semlb{\rhs}{\pp}$. If this does not
hold, then, by Theorem \ref{thm:strongsemantics} the chance is high
that in some reachable $\pp'$ the constraint is violated.  If
$\semub{\lhs}{\pp} \subseteq \semlb{\rhs}{\pp}$ does hold, then we can
apply the following:

\begin{theorem}[Main with Untrusted Roles] 
\label{thm:mainsafety}
Let $\rr = (\gr, \sr)$ be a role monitor. Assume that
$\semub{\lhs}{\pp} \subseteq \semlb{\rhs}{\pp}$. Let $\Sigma$ be a
$\pp$-support of $\semub{\lhs}{\pp}$ for $\rhs$ such that $\Sigma
\subseteq \sr$, and let ${\pp} \longmapsto {\pp'}$ be a (possibly
multistep) change from $\pp$ to $\pp'$. If
\begin{enumerate}
\smallromani
\item $\forall\;\cred \in \pp' \backslash \pp,\ \head(\cred) \not\in \ggr{\lhs}{\pp}{\gr}$, and 
\item $\forall\;\cred \in \pp \backslash \pp',\ \head(\cred) \not\in
  \Sigma$
\end{enumerate}
Then $\semub{\lhs}{\pp'} \subseteq \semlb{\rhs}{\pp'}$.\\[2mm]
\emph{Proof}. Take any  $D \in \semub{\lhs}{\pp'}$,
by Lemma \ref{lem:gpropsafety}, $D \in \semub{\lhs}{\pp}$.  By
assumption, $D \in \semlb{\rhs}{\pp}$, and by Lemma
\ref{lem:spropsafety}, $D \in \semlb{\rhs}{\pp'}$.  Hence the thesis.
\HB
\end{theorem}

Because of Theorem~\ref{thm:mainsafety}, in the presence of untrusted
roles we can deploy a monitoring
procedure very similar to that described after Theorem \ref{thm:main}.
First we check that $\semub{\lhs}{\pp} \subseteq \semlb{\rhs}{\pp}$
holds\footnote{
Even if this does not hold, when neither \lhs\ nor \rhs\ is static,
it is possible that $\semsp{\lhs}{\pp'}
\subseteq \semsp{\rhs}{\pp'}$ for all $\pp'$ reachable from \pp.
However, in general, for the class of constraints we consider,
determining this is \textbf{PSPACE}-hard~\cite{LMW04}, \ie,
intractable.  Thus, our technique makes an efficient 
conservative approximation for the more general constraints we consider.
}.  
While doing this, we compute a \pp-support $\Sigma$ of
$\semub{\lhs}{\pp}$ for $\rhs$---this time a $\Sigma$ such that
$\Sigma \subseteq \sr$.  Second, we have to build
$\ggr{\lhs}{\pp}{\gr}$.  Third, we monitor the roles in
$\Sigma$ and in $\ggr{\lhs}{\pp}{\gr}$ so that each time a statement
defining a role in $\ggr{\lhs}{\pp}{\gr}$ (resp.\ $\Sigma$) is added
to (resp.\ deleted from) $\pp$, the constraint owner is warned.
When the constraint owner receives a
warning, he has to (a) check whether $\semub{\lhs}{\pp} \subseteq
\semlb{\rhs}{\pp}$ still holds, and (b) recompute
$\ggr{\lhs}{\pp}{\gr}$ and $\Sigma$.

\begin{example}
\sloppypar
  Reconsider again Example \ref{ex:HAZMAT}.  Suppose that
  $\mathit{\Emergency.\dept}$ is (the only role) not in $\gr$, then we
  have that $\Emergency.\responsePersonnel \not\in \core{\gr}{\pp}$.
  Therefore
\\[2mm]
$
\begin{array}{lllll}
\multicolumn{5}{l}{\ggr{\Emergency.\hazmatPersonnel}{\pp}{\gr} =}\\
\ \hspace{5mm} \    & \{ & \Emergency.\hazmatPersonnel,\\
           &       & \ATF.\hazmatTraining  & \}
\end{array}$
\\[2mm]
Nonetheless,
if $\ATF.\hazmatDB \in \sr$ we have that
\[\begin{array}{ll}
&\semub{\Emergency.\hazmatPersonnel}{\pp}\\
\subseteq & \semlb{\ATF.\hazmatDB}{\pp}
\end{array}\]
so by Theorem \ref{thm:strongsemantics}
we know that the constraint
\[ \Emergency.\hazmatPersonnel \sqsubseteq \ATF.\hazmatDB\]
is satisfied in all reachable $\pp'$. By Theorem
\ref{thm:mainsafety}, if the two roles 
\Emergency.\hazmatPersonnel, and
 \ATF.\hazmatTraining, prompt a warning when a statement defining one
 of them is added and the role \ATF.\hazmatDB gives a warning when one
 of its statement is removed, then the constraint needs to be
 re-checked only when a warning is given. In that case, we also have
 to recompute $\Sigma$ and
 $\ggr{\Emergency.\hazmatPersonnel}{\pp}{\gr}$.  Theorem
 \ref{thm:mainsafety} guarantees that no matter which changes are made
 to $\pp$, until a warning is given, we still have that every
 reachable\footnote{Notice that changing \pp also changes the
 reachability relation, i.e., the set of reachable $\pp'$s.}  $\pp'$
 satisfies the constraint.
\HB
\end{example}

\section{Related Work}
\label{sec:relatedwork}

In database theory, an integrity constraint is a query that must
remain \emph{true} after the database has been updated. Originally,
integrity constraints were introduces to prevent incorrect updates and
to check the database for integrity. Nevertheless, integrity
constraints have later been used for a number of purposes, ranging
from query optimization to view updating. We refer to
\cite{GGGM98,Das92} for illustrative examples of the uses of integrity
constraints in deductive databases.

In Section~\ref{sec:preliminaries}, we listed several papers
presenting various trust management systems.  None of these
incorporates a notion of integrity constrains.  The work in trust
management that is most closely related is~\cite{LMW04}.  As we
discussed at the beginning of Section~\ref{sec:safety}, that work is
complimentary to ours.  It studies the problem of determining, given a
state \pp, a role monitor \rr, and a constraint $Q$, whether there is
a reachable state in which $Q$ is violated.  By contrast, we analyze
the problem of which roles must have their definitions monitored to
detect when such a \pp is entered.

\section{Conclusion}
\label{sec:conclusion}

We introduce the use, monitoring, and enforcement of integrity
constraints in trust management-style authorization systems.  We
consider the portions of the policy state that must be monitored to
detect violations of integrity constraints.  We also address the extra
difficulty that not all participants in a trust management system can
be trusted to assist in such monitoring, and show how many integrity
constraints can be monitored in a conservative manner so that trusted
participants detect and report if the system enters a policy state
from which evolution in unmonitored portions of the policy could lead
to a constraint violation.

\section{Acknowledgments}
We thank Pieter Hartel and Ha Manh Tran for their precious help and the anonymous referees for their comments.

%
%

\onecolumn
\appendix
\section{Proofs}

\noindent \textbf{Lemma~\ref{lem:gprop}}
Let $\pp' = \pp \u \{\cred\}$, where $\head(\cred) \not\in \g{A.r}{\pp}$, then
\begin{enumerate}
\parentalphi
\item $\semsp{A.r}{\pp}  = \semsp{A.r}{\pp'}$, and 
\item $\g{A.r}{\pp} = \g{A.r}{\pp'}$.
\end{enumerate}
Moreover, if $\pp'$ is obtained from \pp by (a) adding zero or more
statements whose head is not in \g{A.r}{\pp}, and (b) removing zero
or more statements, then
\begin{enumerate}
\setcounter{enumi}{2}
\parentalphi
\item $\semsp{A.r}{\pp} \supseteq \semsp{A.r}{\pp'}$, and 
\item $\g{A.r}{\pp} \supseteq \g{A.r}{\pp'}$.
\end{enumerate}
\emph{Proof}.

(a) Let $P = \SP{\pp}$, and $P' = \SP{\pp'}$. First, summarize some
logic-programming 
notation: we denote by $B_P$ the Herbrand base of $P$ (and $P'$),
consisting of the set of all ground (variable-free) atoms.
$\Ground(P)$
denotes the set of all ground instances of clauses in $P$.
The usual
$\mytp$ operator is defined as follows: let $I \subseteq B_P$, then
$\mytp(I) \ = \ \{ H \ | \ H \lif B_1,\ldots,B_n \in \Ground(P), \mbox{
  and } B_1,\ldots,B_n \in I \}$. As usual, we define $\mytpn{0}(I) :=
I$, and $\mytpn{n+1}(I) := \mytp(\mytpn{n}(I))$. By well-known results
(see e.g., \cite{Apt97}), since $P$ contains no function symbols, for
some $n$ we have that 
\[\mytpn{n}(\emptyset) = M_P = \mbox{ the least Herbrand model of $P$} \]
Now we define the LP-counterpart of $\g{A.r}{\pp}$:
\newcommand{\gatom}{\Gamma_{\mathrm{atom}}}
\newcommand{\gatomcomp}{\overline{\Gamma_{\mathrm{atom}}}}
$\gatom  =  \{ m(B,r,D) \ | \ B.r \in \g{A.r}{\pp} \wedge D \in
\Principals \}$ and the complement
$\gatomcomp  =  \{ m(B,r,D) \ | \ B.r \not\in \g{A.r}{\pp}  \wedge D \in
\Principals\}$.
Furthermore, let $I$ and $I'$ be two sets of ground atoms such that $I' = I \u
\mbox{ some atoms in $\gatomcomp$}$, and $I \subseteq M_P$.  By the
monotonicity of $\mytp$, we have that
\begin{equation}
\label{eq:tpprimebigger}
\mytpp(I')  \supseteq  \mytp(I) 
\end{equation}
We now want to show that 
\begin{equation}
\label{eq:tpprimeingamma}
\mytpp(I') \backslash \mytp(I) \subseteq \gatomcomp
\end{equation}
We proceed by contradiction and assume that there exists $H$ such that 
\begin{equation}
\label{eq:absurd}
H \in \mytpp(I') \backslash \mytp(I) \mbox{ and } H \in \gatom
\end{equation}
Since $H \in \mytpp(I')$, there exists a ground instance $H \lif
B_1,\ldots,B_n$ of a clause $cl \in P$ such that $B_1,\ldots,B_n \in
I'$. Since $H \in \gatom$, $cl \in P$. Therefore $H \in \mytp(I')$.
We now want to show that 
\begin{equation}
\label{eq:nextstep}
B_1,\ldots,B_n \in \gatom 
\end{equation}
Since $I' \backslash I \subset \gatomcomp$, this will demonstrate that
$B_1,\ldots,B_n \in I$, and therefore that $H \in \mytp(I)$, contradicting
(\ref{eq:absurd}).  We distinguish two cases according to the
kind of statement from which $cl$ is generated. Case 1: $cl$ is the
LP-translation of a simple inclusion or intersection inclusion
(not a linking inclusion).
Then $B_1,\ldots,B_n \in \gatom$ by Definition \ref{def:gp}. Case 2:
$cl$ is the LP-translation of a linking inclusion (linked role). Then
$H \lif B_1,\ldots,B_n$ has the form $m(A,r,D) \lif m(A,r_1,B),
m(B,r_2,D)$. By Definition \ref{def:gp}, $m(A,r_1,B) \in \gatom$.
Since $I' \backslash I \subset \gatomcomp$, and $m(A,r_1,B) \in I'$,
we have that $m(A,r_1,B) \in I$.  Since $I \subseteq M_P$, then $B \in
\semsp{A.r}{\pp}$. Therefore, again by Definition \ref{def:gp},
$m(B,r_2,D) \in \gatom$, proving (\ref{eq:nextstep}) (which in turn
contradicts \ref{eq:absurd}).

Now that we have proven (\ref{eq:tpprimeingamma}), since for each $m$ we
have that $\mytpn{m} \subseteq M_P$, from (\ref{eq:tpprimebigger}),
(\ref{eq:tpprimeingamma}) and a straightforward inductive reasoning it
follows that, for each $m$,
\[ \mytppn{m}(\emptyset)  \supseteq  \mytpn{m}(\emptyset)
\mbox{ \ \ and \ \ } 
\mytppn{m}(\emptyset) \backslash \mytpn{m}(\emptyset) \subseteq \gatomcomp
\]
Since the least model of $P'$ and $P$ is the least fixpoint of these
continuous operators on a finite lattice, this demonstrates that
$M_{P'} \backslash M_P \subseteq \gatomcomp$.  Since by definition
$A.r \in \g{A.r}{\pp}$ it follows that $\semsp{A.r}{\pp} =
\semsp{A.r}{\pp'}$. Hence the thesis.

(b) Since $\head(\cred) \not\in \g{A.r}{\pp}$, $\head(\cred)$ is not
reachable from $A.r$.  So removing \cred does not alter the
reachability from $A.r$.

(c) and (d) First notice that, by construction,
\begin{equation}
\label{eq:gdiminishes}
\g{A.r}{\pp} \supseteq \g{A.r}{(\pp \backslash \{cred \})}
\end{equation}
Now, suppose that we have a chain $\pp= \pp_0, \pp_1, \ldots, \pp_n =
\pp_1$, where each $\pp_{i+1}$ is obtained from $\pp_i$ by either
adding a statement whose head is not in \g{A.r}{\pp} or removing a
statement. We now show by induction on $i$ that for each $i \in
[1,n]$: $\semsp{A.r}{\pp} \supseteq \semsp{A.r}{\pp_{i}}$ and
$\g{A.r}{\pp} \supseteq \g{A.r}{\pp_{i}}$, which imply the thesis. The
base case is trivial, as $\pp_1 = \pp$, for the inductive case we have
two subcases: Case 1. If $\pp_{i+1}$ is obtained from $\pp_i$ by
adding a statement \cred such that $\head(\cred) \not\in
\g{A.r}{\pp}$, then by the inductive hypothesis $\head(\cred) \not\in
\g{A.r}{\pp_i}$, and, by statements (a) and (b) we have that
$\semsp{A.r}{\pp_{i}} = \semsp{A.r}{\pp_{i+1}}$ and $\g{A.r}{\pp_{i}}
= \g{A.r}{\pp_{i+1}}$, and the result follows from the inductive
hypothesis.  Case 2. If $\pp_{i+1}$ is obtained from $\pp_i$ by
removing a statement, then the result follows from the monotonicity
of $\semsp{A.r}{\pp_{i}}$ (\ref{eq:monotonicity}), and
(\ref{eq:gdiminishes}).
\HB

\vspace{0.2in}

\noindent\textbf{Proposition~\ref{pro:noncore}}
Let \pp be a set of statements and \gr\ be a set of roles.  If $A.r
\not\in \core{\gr}{\pp}$, then $\semub{A.r}{\pp} = \Principals \u
\{\top \}$.
\\
\emph{Proof}.
\newcommand{\myop}{\ensuremath{\mathit{cl}_{\pp}}}
\newcommand{\myopn}[1]{\ensuremath{\mathit{cl}_{\pp}\!\uparrow{#1}}}
Consider the following closure operator on sets of roles ($\myop:
\wp(\Roles) \ra \wp(\Roles)$). Let $\Delta$ be a set of roles.
\[
\begin{array}{rccl}
\myop(\Delta) & = & & \Delta \\
&& \u & \{ A.r \ | \ A.r \includes B.r \in \pp \mbox{ and } B.r \in \Delta \} \\
&& \u & \{ A.r \ | \ A.r \includes A.r_1.r_2 \in \pp \mbox{ and } A.r_1 \in \Delta \} \\
&& \u & \{ A.r \ | \ A.r \includes A.r_1.r_2 \in \pp \mbox{ and } \exists B \in \semub{A.r_1}{\pp} \mbox{ such that } B.r_2 \in \Delta \} \\
&& \u & \{ A.r \ | \ A.r \includes B_1.r_1 \i \ldots B_n.r_n \in \pp \mbox{ and } \forall i \in [1,n]\ B_i.r_i \in \Delta \}
\end{array}
\]
It is easy to see that $\core{\gr}{\pp}$ is---by
construction---exactly the least fixpoint of $\myop$ containing
$\overline{\gr}$, the
complement of $\gr$. Now, define $\myopn{0}(\Delta) := \Delta$, and
$\myopn{n+1}(\Delta) := \myop(\myopn{n}(\Delta))$. Since $\myop$ is
monotonically increasing, and since $\wp(\Roles)$ is
finite, we have that, for some $n$.
\begin{equation}
\label{eq:fixpoint}
\myopn{n}(\overline{\gr}) = \mbox{ least fixpoint of $\myop$ containing $\overline{\gr}$ } = \core{\gr}{\pp}
\end{equation}
Now, by definition, for every $A.r \in \overline{\gr}$,
$\semub{A.r}{\pp} = \Principals \u \{\top \}$.\\
By the definition of $\myop$, it is straightforward to check that this
implies that for every $A.r \in \myop(\overline{\gr})$,
$\semub{A.r}{\pp} = \Principals \u \{\top \}$.\\
By iterating this reasoning  it is straightforward to check that this
implies that for every $A.r \in \myopn{n}(\overline{\gr})$,
$\semub{A.r}{\pp} = \Principals \u \{\top \}$.\\
The thesis follows from (\ref{eq:fixpoint}). \HB
\vspace{2mm}

\section{Computing the Support Bottom-Up}

We now show how one can compute the support in bottom-up way. We do
this by defining a semantics: $\jssymb: \Roles \ra \wp(\Principals
\times \wp(\Roles))$ for which it holds that if $\js{A.r}{\pp} \ni
\coa{D}{\Sigma}$ then $\Sigma$ is a minimal $\pp$-support of $D$ in $A.r$. 
The construction is parametric wrt the partial order used to define
minimality.

\begin{definition}[Justified Set Semantics $\jssymb$]
\label{def:jss}
In the following algorithm $\cs$ and $\os$ are mappings $\Roles \ra
\wp(\Principals
\times \wp(\Roles) \times {\mathbb N})$. 
We say that
$\cob{D_1}{\Sigma_1}{i_1}$ subsumes $\cob{D_2}{\Sigma_2}{i_2}$ iff
$D_1 = D_2$ and $\Sigma_1 \subseteq \Sigma_2$.
{\small
\begin{tabbing}
xx \= xx \= xx \= xx \= xx \= xx \= xx \= xx \= xx \=\kill
\> \textbf{init phase}\\
\> \> for each role $A.r$, $\cs(A.r) := \emptyset$\\
\> \textbf{repeat}\\
\> \> for each role $A.r$, do $\os(A.r) := \cs(A.r)$\\
\> \> for each $\cred \in \pp$ do\\
\> \> \> if $\cred = A.r \includes B$ then \\
\> \> \> \> \> \emph{remove from \cs(A.r) all triples  subsumed by \cob{B}{\{A.r\}}{1}}\\
\> \> \> \> \> $\cs(A.r) := \cs(A.r) \u \{ \cob{B}{\{A.r\}}{1} \} $\\
\> \> \> if $\cred = A.r \includes B.s$ then \\
\> \> \> \> for each $\cob{D}{\Sigma}{i} \in \cs(B.s)$ do\\
\> \> \> \> \> if $\cob{D}{\Sigma \u \{A.r\}}{i+1}$ is not subsumed by any triple in  $\cs(A.r)$ then \\
\> \> \> \> \> \> \emph{remove from \cs(A.r) all triples  subsumed by \cob{D}{\Sigma \u \{A.r\}}{i+1}}\\
\> \> \> \> \> \> $\cs(A.r) := \cs(A.r) \u \{ \cob{D}{\Sigma \u \{A.r\}}{i+1} \}$\\
\> \> \> if $\cred = A.r \includes A.r_1.r_2$ then \\
\> \> \> \> for each $\cob{B}{\Sigma_1}{i_1} \in \cs(A.r_1)$ do\\
\> \> \> \> \> for each $\cob{D}{\Sigma_2}{i_2} \in \cs(B.r_2)$ do\\
\> \> \> \> \> \> if $ \cob{D}{\Sigma_1 \u \Sigma_2 \u \{A.r\}}{i_1 + i_2}$ is not subsumed by any triple in  $\cs(A.r)$ then \\
\> \> \> \> \> \> \> \emph{remove from \cs(A.r) all triples  subsumed by $ \cob{D}{\Sigma_1 \u \Sigma_2 \u \{A.r\}}{i_1 + i_2}$}\\
\> \> \> \> \> \> \> $\cs(A.r) := \cs(A.r) \u \{ \cob{D}{\Sigma_1 \u \Sigma_2 \u \{A.r\}}{i_1 + i_2} \}$\\
\> \> \> if $\cred = A.r \includes B_1.r_1 \i{} B_2.r_2$ then \\
\> \> \> \> for each $\cob{D}{\Sigma_1}{i_1} \in \cs(B_1.r_1)$ do\\
\> \> \> \> \> if, for some $\Sigma_2$, $i_2$ $\cob{D}{\Sigma_2}{i_2} \in \cs(B_2.r_2)$ then\\
\> \> \> \> \> \> if $ \cob{D}{\Sigma_1 \u \Sigma_2 \u \{A.r\}}{i_1 + i_2}$ is not subsumed by any triple in  $\cs(A.r)$ then \\
\> \> \> \> \> \> \> \emph{remove from \cs(A.r) all triples  subsumed by $ \cob{D}{\Sigma_1 \u \Sigma_2 \u \{A.r\}}{i_1 + i_2}$}\\
\> \> \> \> \> \> \> $\cs(A.r) := \cs(A.r) \u \{ \cob{D}{\Sigma_1 \u \Sigma_2 \u \{A.r\}}{i_1 + i_2} \}$\\
\> \textbf{until} for each role $A.r$, $\os(A.r) = \cs(A.r)$
\end{tabbing}}
Then, for each role $A.r$, we define 
$\js{A.r}{\pp} := \{ \coa{D}{\Sigma}\ | \ \exists i\ \cs(A.r) \ni \cob{D}{\Sigma}{i} \}$. \HB
\end{definition}

The following result demonstrates that this semantics is equivalent to
the standard one, and that it provides us with appropriate
support-sets.

\begin{theorem} Let $A.r$ be a role, $D$ a principal, and $\pp$
  a state. Then $\coa{D}{\Sigma_0} \in \js{A.r}{\pp}$ if and only if
  $\Sigma_0$ is a minimal \pp-support of $D$ in $A.r$.
\end{theorem}
\emph{Proof.}
($\Leftarrow$)  
Assume $\Sigma_0$ is a minimal set of roles such that $D \in
\semsp{A.r}{\pp|_{\Sigma_0}}$.
We show by induction on the construction of
$\mytpsn{\pp|_{\Sigma_0}}{n}(\emptyset)$ that for all $j$ and for each
$A_0.r_0 \in \Sigma_0$, if $m(A_0, r_0,D) \in
\mytpsn{\pp|_{\Sigma_0}}{j}(\emptyset)$, then at some stage in the
execution of the algorithm, for some $i$ and $\Sigma$,
$\cob{D}{\Sigma}{i} \in \cs(A_0.r_0)$ with $\Sigma \subseteq \Sigma_0$.
The desired result then follows by taking $A_0.r_0 = A.r$, by using the
fact, shown below in the second part of the proof, that $\cob{D}{\Sigma}{i}
\in \cs(A_0.r_0)$ implies
$m(A_0,r_0,D) \in 
\mytpsn{\pp|_{\Sigma}}{n}(\emptyset)$, and by using the minimality of
$\Sigma_0$.

\emph{Basis.}  When $j=0$, the result is trivial.

\emph{Step.}  We assume the hypothesis holds for $j$ and show that it
holds for $j+1$.  We proceed by case analysis of the clause used to
add $m(A_0, r_0, D)$ to $\mytpsn{\pp|_{\Sigma_0}}{j+1}(\emptyset)$.  We
show here only the case of linking inclusion; the other cases are similar.

\sloppypar
\emph{Case:} $m(A_0,r_0,?Z) \lif m(A_0, r_1, ?Y), m(?Y, r_2, ?Z)\in
\SP{\pp|_{\Sigma_0}}$.
By definition of \mytp, there exists $B$ such that $m(A_0, r_1, B),
m(B,r_2,D) \in \mytpsn{\pp|_{\Sigma_0}}{j}(\emptyset)$.  So by
induction hypothesis, there exist $i_1, i_2, \Sigma_1, \Sigma_2$ such
that $\Sigma_1, \Sigma_2 \subseteq \Sigma_0$,
$\cob{B}{\Sigma_1}{i_1} \in \cs(A_0.r_1)$, and 
$\cob{D}{\Sigma_2}{i_2} \in \cs(B.r_2)$ by some stage in the
execution.  Consider the first such stage.  In the following
iteration, either $\cs(A_0.r_0)$ already contains a triple that
subsumes $\cob{D}{\Sigma_1 \cup \Sigma_2 \cup \{A_0.r_0\}}{i_1 + i_2
+1}$, or else this triple is added.  In either case, at the end of the
iteration, $\cs(A_0.r_0)$ contains a triple that subsumes
\cob{D}{\Sigma_0}{k}, for all $k$.  (Note $\Sigma_1 \cup \Sigma_2 \cup
\{A_0.r_0\} \subseteq \Sigma_0$.)

($\Rightarrow$)  We show by induction on $i$ that if
$\cob{D}{\Sigma}{i} \in \cs(A.r)$, then $m(A,r,D) \in 
\mytpsn{\pp}{n}(\emptyset)$.
This direction of the theorem then follows because, by the other
direction, all minimal \pp-support are in $\cs(A.r)$, and the
algorithm removes all entries that are subsumed by other entries.

\emph{Basis.} $i = 1$.  In this case, $\Sigma = \{A.r\}$ and there is
a statement $\Cred{A.r}{D} \in \pp$.  In this case $m(A,r,D) \in
\SP{\pp|_{\Sigma}}$, so $m(A,r,D) \in \mytpsn{\pp}{j+1}(\emptyset)$ for all
$j \in {\mathbb N}$.

\emph{Step.} We assume the hypothesis holds for all $i \leq k$ and
show that it holds for $i=k+1$.  We proceed by case analysis of the
statement used to add \cob{D}{\Sigma}{k+1} to $\cs(A.r)$.  We show
here only the case of linking inclusion; the other cases are similar.

\emph{Case:} \Cred{A.r}{A.r_1.r_2}.  In this case there are $\Sigma_1$, $\Sigma_2$,
$i_1$, $i_2$, and $B$ such that $\cob{B}{\Sigma_1}{i_1} \in
\cs(A.r_1)$, $\cob{D}{\Sigma_2}{i_2} \in \cs(B.r_2)$, $k = i_1 + i_2$,
and $\Sigma = \Sigma_1 \cup \Sigma_2 \cup \{A.r\}$.
By induction hypothesis, $m(A,r_1, B) \in
\mytpsn{\pp|_{\Sigma_1}}{n}(\emptyset)$ and 
$m(B,r_2, D) \in \mytpsn{\pp|_{\Sigma_2}}{n}(\emptyset)$.
By monotonicity of \mytp\ in $P$, it follows that $m(A,r_1, B),
m(B,r_2, D) \in \mytpsn{\pp|_{\Sigma}}{n}(\emptyset)$.
Consider the first $j$ such that 
$m(A,r_1, B), m(B,r_2, D) \in \mytpsn{\pp|_{\Sigma}}{j}(\emptyset)$.
Because $m(A,r,?Z) \lif m(A,r_1,?Y),m(?Y,r_2,?Z)$ is in
\SP{\pp|_{\Sigma}}, it follows that $m(A,r_1,D) \in
\mytpsn{\pp|_{\Sigma}}{j+1}(\emptyset)$, the latter being a subset of 
$\mytpsn{\pp|_{\Sigma}}{n}(\emptyset)$.
\HB

It must be acknowledged that the algorithm given here may construct a
value for \cs\ whose size is combinatorial in the size of \pp.  In
practice, a variant of this algorithm should be used in which a small
constant number of entries in $\cs(A.r)$ are stored for each $D \in
\semsp{A.r}{\pp}$.
 
\end{document}